# Deep Vision-Inspired Bubble Dynamics on Hybrid Nanowires with Dual Wettability

Jonggyu Lee[1,§], Youngjoon Suh[1,§], Max Kuciej[2,3], Peter Simadiris[1], Michael T. Barako[3], Yoonjin Won[1,4,*]

Co-First Authors[§], and Corresponding Author[*]

[1]Department of Mechanical and Aerospace Engineering, University of California, Irvine, Irvine, CA, 92697, USA.

[2]Departement of Material Science and Engineering, University of California, Los Angeles, Los Angeles, CA, 90095, USA

[3] NG Next, Northrop Grumman Corporation, Redondo Beach, CA, 90278, USA.

[4] Department of Electrical Engineering and Computer Science, University of California, Irvine, Irvine, CA, 92697, USA

Corresponding Author Email: won@uci.edu

## ABSTRACT

The boiling efficacy is intrinsically tethered to trade-offs between the desire for bubble nucleation and necessity of vapor removal. The solution to these competing demands requires the separation of bubble activity and liquid delivery, often achieved through surface engineering. In this study, we independently engineer bubble nucleation and departure mechanisms through the design of heterogeneous and segmented nanowires with dual wettability with the aim of pushing the limit of structure-enhanced boiling heat transfer performances. The demonstration of separating liquid and vapor pathways outperforms state-of-the-art hierarchical nanowires, in particular, at low heat flux regimes while maintaining equal performances at high heat fluxes. A machine vision-based framework realizes the autonomous curation and extraction of hidden big data along with bubble dynamics. The combined efforts of materials design, deep learning techniques, and data-driven approach shed light on the mechanistic relationship between vapor/liquid pathways, bubble statistics, and phase change performance.

## INTRODUCTION

Heat transfer surfaces for boiling exhibit a tradeoff between achieving high boiling efficacy and upper boiling limits.[1] The boiling heat transfer efficacy is typically quantified by the heat transfer coefficient (HTC), which relates the superheat temperature to the bubble activity-induced interfacial heat flow, and the critical heat flux (CHF), which establishes the upper limit of boiling heat flux as limited by liquid delivery.[2-10] To perform well on both aspects an ideal boiling surface must reconcile the competing demands for vapor production/extraction and liquid transport.[11]

According to classical nucleation theory, traditional homogeneous wicks have often been construed to be limited to either high-heat flux applications or low-heat flux, high-performance applications because they must compromise between bubble nucleation and departure.[12, 13] Since bubble nucleation is governed by the interfacial activation energy barriers required for bubbles to form on the heated surface,[12, 13] surfaces that favor bubble nucleation typically possess high surface roughness and intrinsic wettability.[8, 9, 12-14] However, these surface properties can have negative effects on vapor removal, hence limiting liquid delivery to the heated surface.[15] For example, past work shows that rough hydrophobic surfaces exhibit high HTC at low heat flux regions, but are plagued with low CHF due to retarded contact line movement and bubble overpopulation.[16] On the other hand, hydrophilic surfaces or wicking surfaces show superior CHF by preventing dry-out of the heated surface at high heat flux regions, but suffer from low HTC at low heat flux regions due to the large activation energy barriers for bubble nucleation.

In response to the boiling dilemma, modern boiling surfaces employ heterogeneous surface properties as a multifaceted approach to ensure favorable bubble dynamics and efficient liquid/vapor pathways.[17-22] A common research thrust is to utilize hydrophilic, micro/nano hierarchically enhanced surfaces with two or even multiple length scales.[23-25] These multi-tier-roughness surfaces achieve some of the highest reported boiling performances, which is attributed to the activation of more nucleation sites, effective evaporative surface area increase, and capillary-assisted surface rewetting enhancement.[1, 25-27] Another effective approach recently receiving attention is the use of surfaces with opposing wettability.[15, 23, 28, 29] This method capitalizes on the juxtaposition of high and low activation energies to promote bubble nucleation on designated spots while continuously replenishing the rest of the surface with liquid.[15, 30] To date, spatially mixed wettabilities have been realized by installing hydrophobic islands on hydrophilic surfaces,[15] or through stochastic deposition techniques,[31] which, in turn, have limited surface wetting areas.[23] Therefore, this study pursues the rational question of whether the advantages of hierarchical structures and biphilic surface properties can be combined and fine-tuned to maximize boiling heat transfer performances. To answer this question, it is imperative to build a strong mechanistic connection of surface property effects on boiling efficacy.[32] The mechanistic investigation of the highly intertwined relationship between surface properties and boiling heat and mass transfer, in turn, requires a rigorous analysis of bubble dynamics.

To quantify the collective bubble evolution behavior crucial for understanding boiling heat transfer mechanisms, we develop a vision-based framework that extracts *in situ* bubble features at single-bubble resolutions. Typically, a boiling surface emits hundreds of bubbles per second that individually exhibit complex movements, interactions, and morphologies, even near the onset of nucleate boiling (ONB), which poses a challenge for researchers to consistently annotate and track them.[32] Furthermore, high-level boiling features such as departure bubble diameter, departure frequency, and nucleation site density require connecting individual features throughout time. Therefore, an optical measurement taken at 2000 fps could easily translate to ~$10^5$ features per second. That is, the sheer volume of information from a single experiment makes manual bubble analysis methods difficult to capture representative bubble dynamic trends. Our deep learning framework, further referred to as Vision-Inspired Online Nuclei Tracker (VISIONiT), is customized to recognize bubble instances in every frame and to autonomously extract rich bubble features from massive datasets that adequately represent the boiling system.[32, 33]

Motivated by the promise of heterogeneous structures and empowered by the new vision-based characterization method, here we suggest using vertically aligned hierarchical nanowire (NW) arrays as boiling wicks, where each NW contains biphilic segments. The base segment is composed of a previously-published superhydrophilic CuO NW array,[26] which expeditiously draws liquid to the heated surface and facilitates bubble departure. The top segment is comprised of a hydrophobic Ni NW array that lowers the local activation energy barriers, thereby promoting bubble nucleation and triggering faster system transitions to the more efficient nucleate boiling region. The NW segments are vertically juxtaposed to fully separate vapor/liquid pathways, thereby segregating bubble activity and liquid transport functions for maximized HTC and CHF. We characterize the liquid transport and corresponding boiling heat transfer, which depends on the NW compound ratio and order. We then correlate the boiling performances with physical features that are extracted by VISIONiT. Our work not only reveals the undiscovered impact of heterogeneous wettability on pool boiling heat transfer, but also offers characterization methods that realize the autonomous curation and extraction of hidden big data residing within visual imagery.

## RESULTS

**Segmented Nanowire Concept and Design**

To better design boiling curves, we propose a segmented NW design where the region adjacent to the interface is superhydrophilic whereas the region at the top of the surface is rough and hydrophobic, thus segregating the optimal properties to their respective regions (**Figure 1**). The segment materials (Cu and Ni) are selected based on thermal conductivity (398 $Wm^{-1}K^{-1}$ for Cu and 106 $Wm^{-1}K^{-1}$ for Ni), intrinsic contact angle (~68° for Cu and ~101° for Ni), availability for electrochemical deposition, and corrosion resistance. The NWs are rendered biphilic by decorating the Cu portion with superhydrophibic CuO nanofeatures via selective oxidation. Unlike most spatially patterned biphilic boiling surfaces, our segmented NW design maximizes the projected surface area for liquid drawing and bubble nucleation by vertically aligning portions with opposing wettabilities. The hydrophobic Ni NWs nucleate bubbles at higher densities than the superhydrophilic CuO NWs at lower superheats due to low bubble activation energy and high density of entrapped vapor sites (Figure 1a). At the same time, the large contact angle of the Ni NWs hinders bubble departure such that neighboring nucleation sites are forced to merge before departure. In contrast, the hydrophilic CuO NWs have slightly delayed ONB points, but quickly release bubbles from the surface once they form. Furthermore, liquid is more easily replenished to the heated surface through the water-friendly CuO NWs, compared to the Ni NWs (Figure 1b). We postulate that our proposed design will combine the advantages of both Ni and CuO NWs, thereby possessing both favorable bubble dynamics and higher dry-out thresholds (Figure 1c).

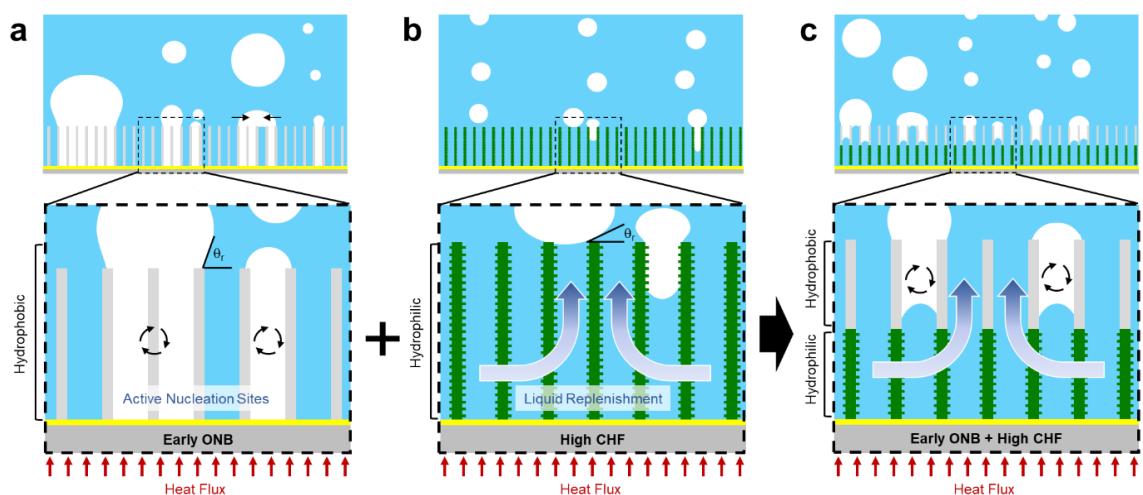

**Figure 1.** Schematic of pool boiling mechanisms using segmented nanowires (NWs). The illustrations depict bubble nucleation/departure and liquid rewetting behaviors for various NW cases. (a) Hydrophobic NWs have low activation energy barriers, meaning that they are generally proficient at nucleating bubbles. Furthermore, the large rewetting angles allow for vapor entrapment in surface cavities, which become active nucleation sites. However, bubble departure is delayed due to retarded contact line motions, causing bubbles to merge before departure. (b) Hydrophilic NWs enable efficient liquid replenishment to the surface and delay the formation of resistive vapor blankets that lead to the surface dry-out. In contrast to hydrophobic surfaces, bubble nucleation is deterred due to high activation energies but are easily detached from the surface once they form. (c) The proposed segmented NW design aims to capitalize on the advantages of both low activation energy barriers and high liquid delivery through dual wettability.

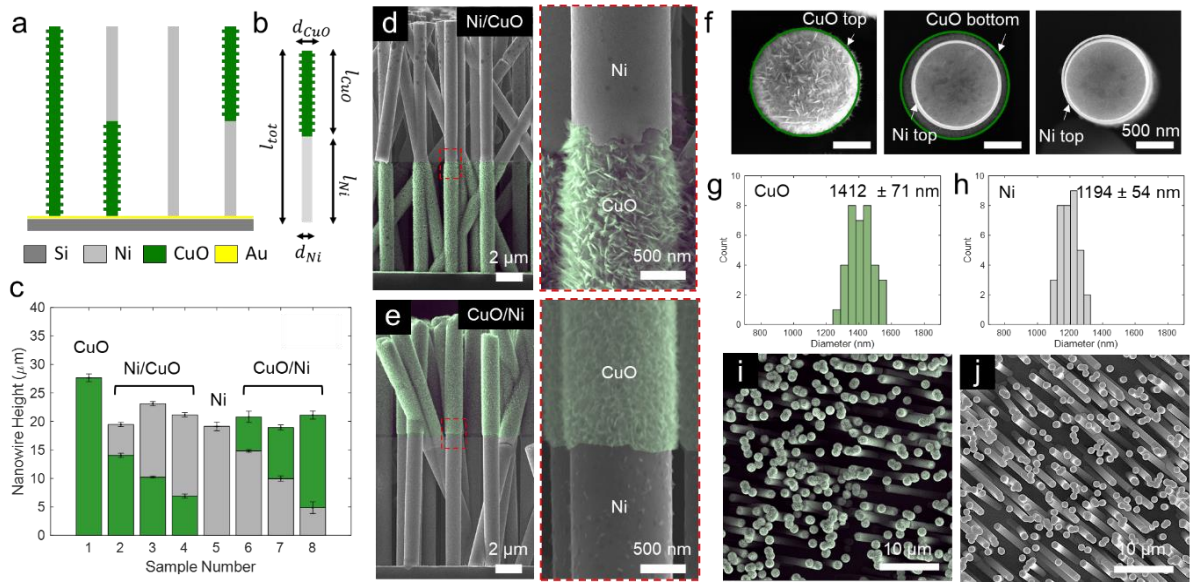

**Figure 2.** Nanoscale morphologies of segmented NWs. (a) We classify our NWs into four categories and denote them based on their material and composition order. The figure illustrates CuO, Ni/CuO, Ni, and CuO/Ni NWs from left to right, respectively. (b) The NWs are further described by quantifying the CuO content, represented as CuO portion length to total length ratio $l_{CuO}/l_{tot}$, where $l_{tot} \approx 20\ \mu m$ for all NWs in this study. (c) The chart lists eight NW samples with varying composition ratios and order. The green and grey color represents CuO and Ni portion, respectively. Side-view scanning electron microscopy (SEM) images show the nanoscale morphology of segmented (d) Ni/CuO NWs (sample 3) and (e) CuO/Ni NWs (sample 7). (f) SEM images show the CuO/Ni, Ni/CuO, and Ni NWs, respectively. The top-view images clearly show selective diameter growth in portions where CuO nanofeatures are grown. The histograms show the distribution of nanowires' diameter of the (g) CuO portion and (h) Ni portion. SEM images of the wide top-view of the (i) CuO/Ni and (i) Ni/CuO NWs confirm minimal difference in terms of tortuosity, density, and orientation.

To experimentally reconcile our hypothesis, we fabricate a series of NWs with varying composition ratios and orders as shown in **Figure 2**a-c. Singular NWs are denoted using their primary materials (e.g., Ni NWs, CuO NWs), whereas segmented NWs are additionally labeled by the portion order. For example, CuO/Ni indicates that the CuO compound is on the top, while the Ni element is on the bottom portion of the segmented NWs, and vice versa (Figure 2c). The NWs are prepared via electrodeposition techniques using porous sacrificial templates (see Methods Section for details).[34, 35] Based on our previous work on hierarchical CuO NWs,[26] we select an optimal template pore diameter of 1000 $nm$ that renders NWs with high permeability and capillary pressure. By tuning the electrodeposition time and element order, we fabricate NWs with near-constant length ($l_{tot} \approx 21 \pm 2\ \mu m$) but having varying composition ratios $l_{CuO}/l_{tot}$ (see Figure 2b and c and **Table 1** to see the full list of the NWs).

**Table 1.** Summary of sample details for segmented nanowires after oxidation.

| Sample Number | Segment Type | $l_{cu}/l_{tot}$ | $\varphi$ | $\theta_{app}$ (°) |
|---|---|---|---|---|
| Sample 1 | CuO | 1.00 | 0.64 | 23.7 |
| Sample 2 | Ni/CuO | 0.72 | 0.67 | 60.4 |
| Sample 3 | Ni/CuO | 0.44 | 0.70 | 55.8 |
| Sample 4 | Ni/CuO | 0.33 | 0.71 | 48.3 |
| Sample 5 | Ni | 0 | 0.74 | 103.5 |
| Sample 6 | CuO/Ni | 0.29 | 0.71 | 25.7 |
| Sample 7 | CuO/Ni | 0.47 | 0.69 | 26.6 |
| Sample 8 | CuO/Ni | 0.84 | 0.66 | 25.8 |

**Nanoscale Morphology**
To introduce additional surface roughness, secondary CuO nanofeatures are selectively grown on the as-fabricated Cu portion of the NWs via chemical immersion (Figure 2d and e). The targeted oxidation onto the Cu NWs produces dense, sharp CuO nanoscale features that selectively increase the local Cu NW diameters as shown in Figure 2f-h. SEM images (Figure 2f) clearly show that the Ni portion retains a smooth surface with negligible morphological changes, whereas the roughness and effective diameter of the CuO portion increases. It should be noted that Ni is corrosion resistant and is thereby unaffected by the chemical oxidation process. The segmented NW diameters are measured from side-view images (see Figure S1 for details), where the distribution is shown in Figure 2g and h. The average Ni NW and CuO NW diameter is measured as $d_{Ni} = 1194\ nm$ and $d_{CuO} = 1412\ nm$, respectively. Since the chemical oxidation has no noticeable effects on the NW's natural tortuosity (Figure 2i and j), the porosity becomes a function of the effective NW diameter, which differs for each segment. By factoring in the diameter differences between segments into porosity calculations, we report a porosity of $0.6 \pm 0.05$ across all NW samples. Please see Supporting Information S1 for porosity calculation. We additionally report that the electrodeposition order decides the base material surface properties as well. To fabricate Ni/CuO NWs, the Cu material is deposited first, generating a very thin layer of Cu between the Au base and template. During the selective oxidation process, this thin layer becomes oxidized as well, reinforcing a superhydrophilic CuO base layer of ~ $400\ nm$ as shown in Figure S1. By contrast, the Ni and CuO/Ni NW base do not show any level of oxidation.[36]

**Nanostructure-Coupled Surface-Liquid Interactions**
We investigate the intrinsic wettability effect on interfacial phenomena by performing a contact angle test on all surfaces (**Figure 3**a). In comparison to the plain surface tests in Figure S2, Figure S3 shows that the NWs containing Cu content significantly increase in hydrophilicity after chemical oxidation ($60\% < \Delta\theta < 80\%$), regardless of the composition order. In addition, Ni/CuO NW surface exhibits less severe contact angle drops compared to CuO/Ni and CuO NW surfaces (Figure S3), maintaining contact angles over 50° across all composition ratios. We note that the errors between different composition ratios all lie within one standard deviation (Figure 3a). These findings not only highlight the extreme wettability bandwidth that hierarchical NWs possess, but also imply that the interfacial interactions are, to some degree, governed by the initial liquid-solid contact material.

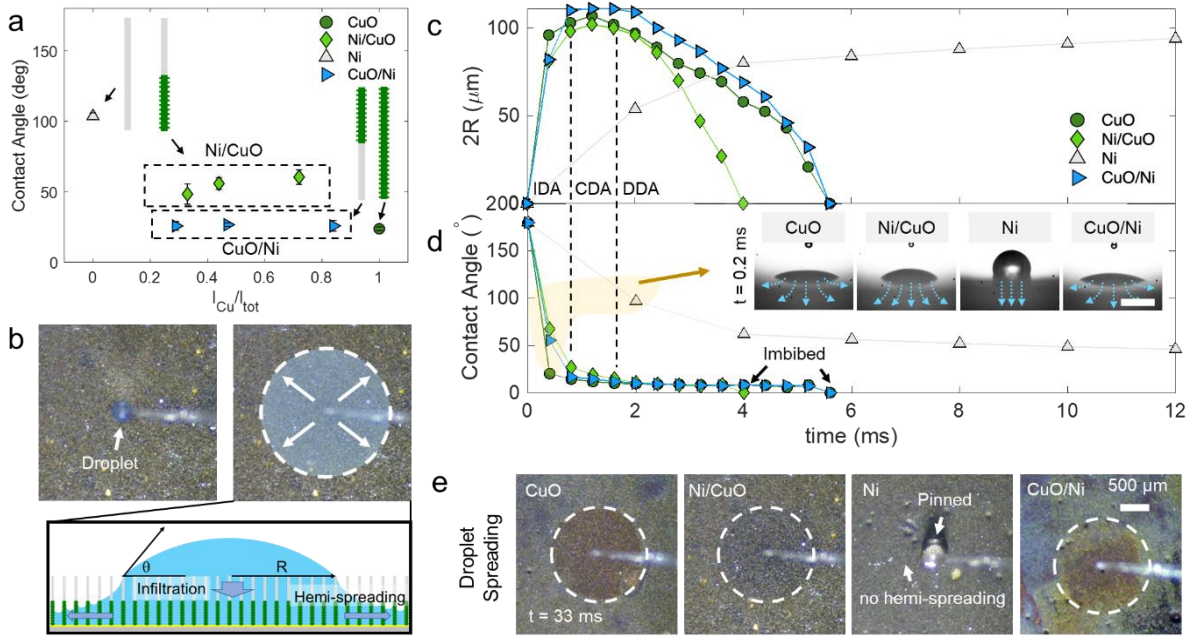

**Figure 3.** Surface wetting characteristics of NWs. (a) The contact angles of segmented NWs are evaluated with the sessile drop method. (b) Top-view optical images of a droplet spreading during the sessile drop test. The illustration shows the hemi-spreading of a droplet on top of a Ni/CuO NW array. The plots for the dynamic (c) diameter of the drawing area and (d) contact angle describe the full droplet imbibition process for representative NW cases, in which the segmented NW cases have $l_{CuO}/l_{tot} \approx 0.5$. Three distinct IDA, CDA, and DDA phases characterize the droplet imbibition. Contact angles are measured near the end of the CDA phase. The inset shows droplets on different samples in the highlighted regions, where liquid delivery pathways are illustrated in blue arrows. The scale bar represents ~170 $\mu m$. (e) The images show representative images of droplets spreading at t = 33 $ms$. The droplets on segmented NWs with CuO contents are fully imbibed, where spread marks are circled in white. The droplets on the Ni NWs, on the other hand, are pinned to the Au sample base. Light refraction-induced color changes can be used to characterize which portion of the NWs spreading occurs.

To study how material composition order influences the wetting behavior, we observe *droplet spreading* via top view imaging. Once a sessile drop is placed on the NW array, the droplet is infiltrated into the pores and spreads via capillary actions in the radial direction as shown in Figure 3b. This behavior is also known as "hemi-spreading" [37, 38] and results from Cassie to Wenzel state transitions.[39, 40] The hemi-spreading occurs when measured contact angles are smaller than the critical contact angle $\theta_c = cos^{-1}(1-\varepsilon)/(r-\varepsilon)$, where $r$ is the roughness. In this computation, we assume NW arrays are aligned in hexagonal arrays, which gives $r = 1 + 2\pi d l_{tot}/\sqrt{3}p^2$, where the pitch between NWs is $p = (2/\sqrt{3}N)^{0.5}$. Accordingly, the critical contact angle is calculated as $\theta_c = 87°$ for $r = 21$ and $p = 2240\ nm$, forecasting that hemi-spreading will occur for all nanowires with CuO components. Our experimental results agree well with the theoretical predictions, showing hemi-spreading behavior for all samples except for the Ni NW ($\theta > \theta_c$) sample (see Figure S4 for top-view spreading images). Interestingly, we observe distinctive two-step spreading behaviors, characterized by light refraction-induced color changes during spreading events (Figure S4). The NW samples where CuO is the initial contact material to the falling droplet (Samples 1, 6-8) display high-contrast color changes, whereas the Ni/CuO samples (Samples 2-4) show little to no color changes. This implies that the droplet spreads immediately after liquid-solid contact on CuO-top NWs while following a vertical infiltration-to-spreading route for Ni-top NWs as illustrated in Figure 3b. The sessile droplet retains more of its spherical shape when they meet hydrophobic Ni NWs, which explains why Ni/CuO NWs possess an overall higher contact angle than CuO/Ni NWs (Figure 3a).

To verify that the microscopic liquid transport through the hierarchical CuO compounds governs interfacial phenomena, we observe additional *droplet spreading* via side-view imaging. Droplet spreading and imbibition into porous media can be analyzed by a classical liquid infiltration theory that shows three phases of droplet absorption.[36, 41-43] The three phases of the imbibition are described as the increasing drawing area (IDA), constant drawing area (CDA), and decreasing drawing area (DDA). Each phase is categorized by temporal changes in the diameter (=2R) of the drawing area and dynamic contact angles (Figure 3c and d). Immediately after initiating liquid-solid contact, the droplet undergoes the IDA phase, where 2R increases while the contact angle decreases until it assumes the advancing contact angle.[44] In the sequential CDA phase, the droplet is absorbed into the porous surface at a relatively constant 2R and decreasing contact angle. Once the contact angle assumes the receding contact angle, the droplet is fully consumed (i.e., imbibed) into the media in the DDA phase, decreasing both 2R and contact angle in the process. The segmented NWs with CuO components show drastically different absorption characteristics compared to the Ni NWs (Figure 3c and d). While all other NWs display rapid transitions through the IDA, CDA, and DDA phases, the droplet on the Ni NW remains pinned to the surface for an extensive time period (~500 $ms$), suggesting an innately high spreading resistance in the radial direction (Figure 3e). Droplets are imbibed into Ni/CuO NWs at the quickest rate (~4 $ms$) despite them being more hydrophobic than CuO and CuO/Ni NWs (Figure 3a). We speculate that the hydrophobicity of Ni NWs at the top direct more liquid towards the superhydrophilic CuO base (Figure 3d inset) to enhance the wicking effects (Figure S1).

**Capillary-Induced Wicking Performance**

As the next step, we quantify the capillary-assisted *liquid transport performance* of our segmented NWs by performing a liquid rate-of-rise test. (see Methods Section, and Figure S5). In brief, when a vertically oriented sample contacts the surface of a liquid pool, a meniscus forms and propagates through the porous media via capillary action (**Figure 4**a). The capillary performance parameter defined by permeability over effective pore radius ($K/R_{eff}$) is obtained from the Lucas-Washburn equation:[33,34,60]

$$h^2 = \frac{4\sigma}{\phi\mu}\frac{K}{R_{eff}}t \qquad (1)$$

where $h$ is the liquid height, $t$ is the time, $\mu$ is the liquid viscosity, and $\sigma$ is the surface tension of the liquid. The effective pore radius is defined as $R_{eff} = R_p/\cos\theta_s$, where $R_p$ is the pore radius and $\theta_s$ is the static contact angle. As expected, the Ni NWs show very low $K/R_{eff}$ values (Figure 4a), which confirms that the Ni NWs have minimal contributions to capillary-induced liquid transport (Figure S6). Furthermore, we observe a trendline that indicates that the $K/R_{eff}$ values of hybrid NWs (i.e., Ni/CuO and CuO/Ni NWs) converge towards the pure CuO NW sample value as the CuO compound ratio increases (Figure 4b). This phenomenon can mechanistically be explained by primary and secondary wicking motions through the segmented NWs.

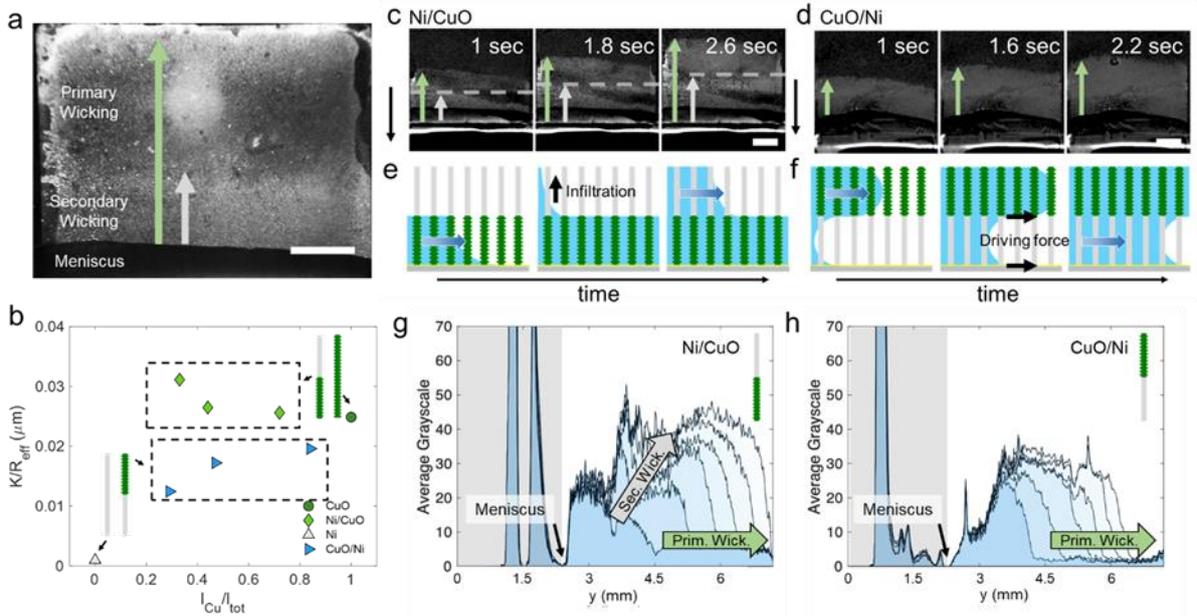

**Figure 4.** Capillary-driven wicking performance of NWs. (a) Segmented NWs exhibit unique two-step capillary wicking behaviors with primary and secondary wicking fronts. Liquid rise is captured by a microscopic optical camera at 60 fps. The scalebar is 2 mm. (b) The capillary performance parameter ($K/R_{eff}$) of the NWs converge towards the CuO NW value as the CuO content increases. Timelapse (c) optical images and (e) illustrations of Ni/CuO NW wicking. The secondary wicking through the Ni NW is primarily due to in-plane liquid infiltration mechanisms. Gray dashed lines indicate the secondary wicking front. Timelapse (d) optical images and (f) illustrations of CuO/Ni NW wicking. The infiltration-induced secondary wicking is assisted by the hydrophilic Au base. The scale bars represent 2 mm. (g) and (h) Changes in the average image grayscale relative to the first frame can visualize primary and secondary wicking. The histogram color fades as time increases. Secondary wicking is identified by increases in the light intensity, which is shown in the plot as vertical upwards histogram shifts.

As evident in the spreading behaviors (Figure 3), the segmented NWs possess a unique *heterogenous wettability*, where the intrinsically opposing wetting characteristics of the two compounds are drastically intensified through nanofeatures. As a result, the liquid wicks through the segmented NWs in primary and secondary phases as shown in Figure 4c-f. The primary and secondary wicking fronts can be identified through distinctive surface light refraction intensities emitted from the liquid at differing in-plane depths. The primary and secondary wick propagations are quantified through an average grayscale analysis, where pixelwise intensities (i.e., light refraction intensities) are used to locate the primary and secondary wicking fronts (Figure 4g and h). While all segmented NWs draw liquid in a two-step process, the two-step wicking is only observable in Ni/CuO NWs, owing to the secondary wicking front's position relative to the imaging direction (Figure 4e; See Movie S1 for real-time visualization of the two-step wicking process). Within the Ni/CuO NWs structures, the secondary wicking is induced by in-plane infiltration mechanisms through the Ni NWs, as illustrated in Figure 4e and f. The primary wicking occurs in the superhydrophilic CuO NW region and is dictated by microscopic liquid pathways with the smallest effective wicking distance (Figure S7). By reviewing the accrued capillary performance results in Figure 4b, we speculate that maximum $K/R_{eff}$ is achieved at the superhydrophilic thin CuO base layer shown

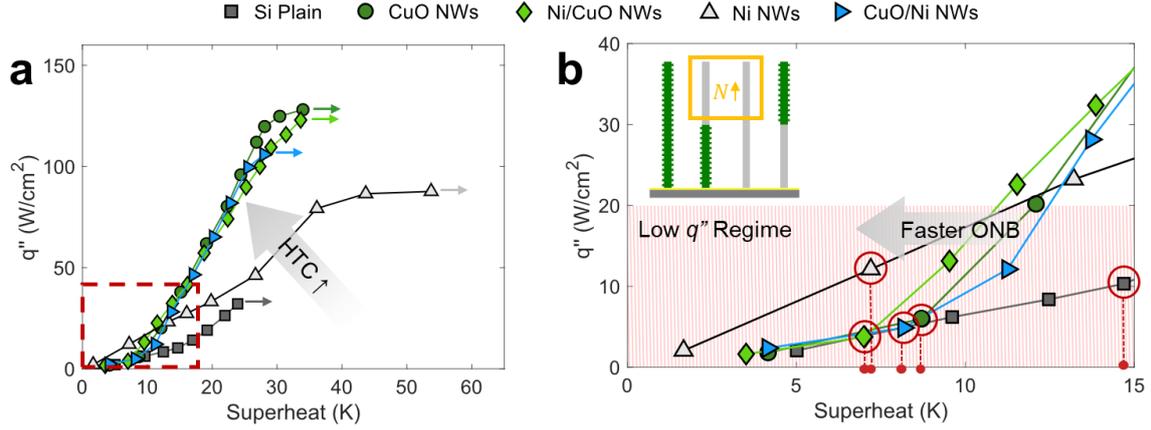

**Figure 5.** Pool boiling heat characterization on the structured surfaces. (a) Boiling heat flux ($q''$) as a function of wall superheat ($\Delta T = T_s - T_{sat}$). The critical heat flux (CHF) is marked with arrows. The boiling curve slope characterizes the heat transfer coefficient (HTC). The red box represents the (b) zoomed in boiling curve at the low $q''$ regime. The onset of nucleate boiling (ONB) is circled in red and marked in the $x$-axis for clarity. The inset illustrates the effectiveness of Ni-top NWs at facilitating bubble nucleation and increasing bubble numbers.

in Figure S1. Since out-of-plane liquid deliver primarily takes place in the CuO portion, Ni/CuO NWs have relatively better capillary performances compared to CuO/Ni NWs (Figure 4b) due to combined wicking effects from the CuO NWs and the CuO base. We accordingly attribute the maximum $K/R_{eff}$ value for Ni/CuO NWs to CuO base-dominated capillarity at low CuO contents, where this dominance fades as the hydraulic resistance relatively increases with higher $l_{cu}/l_{tot}$ ratios. In contrast to Ni/CuO NWs, CuO/Ni NWs lacks a superhydrophilic base and $K/R_{eff}$ scales as a function of CuO content for CuO/Ni NWs as available CuO liquid transport pathways increase. We emphasize that the reported $K/R_{eff}$ values ($0.01 - 0.03\ \mu m$) for segmented NWs is approximately 10× higher than our recently reported high-performance hierarchical nanoscale wick designs.[45, 46]

**Pool Boiling Two-Phase Heat Transfer**
Having characterized the surface properties of our hierarchical nanostructures, we proceed to quantify the *pool boiling heat transfer* characteristics. All experiments are conducted on a custom pool boiling setup presented in Figure S8 and Methods Section. The boiling properties are measured by the steady-state surface heat flux $q''$ with respect to superheat $\Delta T = T_s - T_{sat}$, at incrementally increased input heat loads. In the resulting boiling curve (**Figure 5**a), the CHF is identified as the local maximum, and the HTC = $q''/\Delta T$ is represented by the boiling curve slope.

All NW surfaces display significant HTC and CHF improvements, regardless of the material and composition, in comparison to an unmodified Si surface. A summary of the experimentally measured boiling performances is provided in **Table 2**. Considering the poor capillary performance of the pure Ni NW, even this surface had a CHF of 79 $Wcm^{-2}$, which is a ~150% increase than the unmodified surface, which we attribute to in-plane infiltration-induced liquid delivery mechanisms described in previous sections (Figure 3 and Figure 4). By implementing the superhydrophilic, hierarchical nanofeatures originated from CuO NWs, we manipulate the curve to achieve even higher CHFs of 105 – 128 $Wcm^{-2}$, where the enhancement ratio compared to the unmodified surface approaches the theoretical limit of 4 mentioned in previous literatures (Figure 5a).[25, 47] Simultaneously, the boiling-curve-averaged HTC, $\bar{h} = \frac{1}{\Delta T_{CHF-ONB}} \int_{\Delta T_{ONB}}^{\Delta T_{CHF}} h(\Delta T) \, d(\Delta T)$, shows a 103% and ~212% increase for Ni NWs and NWs with CuO content, respectively, compared to the unmodified Si surface. However, a closer inspection reveals an opposite trend at low heat fluxes ($q''_{low} < 20 \; Wcm^{-2}$), where the Ni NW surface achieves the highest $q''_{low}$ region HTC, $\bar{h}_{low} = \frac{1}{\Delta T_{q''_{low}-ONB}} \int_{\Delta T_{ONB}}^{\Delta T_{q''_{low}}} h(\Delta T) \, d(\Delta T)$, enhancement (~2.3× higher) compared to the unmodified Si surface (Figure 5b). Moreover, the NWs with Ni content show advanced boiling incipience (i.e., ONB), which implies that these surfaces can operate under the more efficient nucleate boiling regime at earlier stages. As expected, the Ni/CuO NW surface proposed to take advantages of both Ni and CuO contents performs exceedingly well on both ends of the boiling curve, with a 207% increase of $\bar{h}$, a 70% increase of $\bar{h}_{low}$, a 293% increase of CHF, and a 52% earlier ONB compared to the control Si surface.

**Table 2.** Summary of experimentally measured boiling heat transfer performances.

| Surface Type | $\bar{h}$ (Wcm⁻²K⁻¹) | $\bar{h}_{low}$ (Wcm⁻²K⁻¹) | CHF (Wcm⁻²) | $\chi$ | $\psi$ |
|---|---|---|---|---|---|
| Si Plain | 1.09 | 0.74 | 32.1 | 1 | 1 |
| CuO NWs | 3.47 | 1.14 | 128.08 | 1.54 | 3.9 |
| Ni/CuO NWs | 3 | 1.23 | 122.93 | 1.66 | 3.82 |
| Ni NWs | 1.73 | 1.71 | 79.25 | 2.30 | 2.46 |
| CuO/Ni NWs | 2.39 | 1.01 | 105.93 | 1.37 | 3.3 |

The further demonstration of a *deep learning-based computer vision framework* (i.e., VISIONiT) enables us to study how the stunning variety of collective structure-dependent bubble nucleation behaviors regulate the boiling performances. VISIONiT chiefly employs object detection, object tracking, and data processing (Figure S9a), to autonomously curate >100,000 rich, physical descriptors per dataset from boiling images at single-bubble resolution. The object detection module identifies and labels every bubble per image with instance-specific pixel-wise masks, where each mask is assigned a unique identifier (ID). Then, the IDed masks are linked together with respect to time through k-dimensional (k-d) tree algorithms in an object tracking module. The tracked spatiotemporal features are finally post-processed to extract features such as the departure bubble diameter $D_b$, the bubble departure frequency $f_b$, and nucleation site density $N_b$. A more detailed description of the framework can be found in Supporting Information S3 and previous work.[33] The model performance and training procedures are documented in the Methods Section.

According to mechanistic pool boiling models,[48-56] $D_b$, $f_b$, and $N_b$, are key parameters used to partition the heat flux into natural convection $q''_{nc}$, evaporation $q''_{ev}$, and bubble departure-driven forced convection $q''_{fc}$ heat transfer modes, see Supporting Information Section S3.2. and Equation S14-17. The departure bubble diameter $D_b$ is indicative of the vapor production rate and is governed by buoyant and surface tension forces acting on the bubble.[57] When $D_b$ is small, boiling heat transfer is determined by a balance between $f_b$ and $N_b$. For example, the heat flux of a surface with high $f_b$ but low $N_b$ may be governed by $q''_{nc}$ due to the large effective non-bubble-influenced area $A_{nbi} = (1 - N_b \pi D_b^2/4)$ exposed to natural convection. When $D_b$ is large, $q''_{nc}$ naturally reduces and the heat transfer is gradually dictated by $q''_{ev}$ as it scales with ~$D_b^3$. Therefore, it is imperative to acquire $D_b$, $f_b$, and $N_b$ to understand the underlying mechanisms that govern boiling heat transfer in the proposed NW designs. See Supporting Information S3 for bubble detection algorithm details and Table S1 for extracted bubble parameters.

Past studies have shown evidence that the bubble statistics at the ONB can serve as good indicators of the CHF because it is a representative stage where the heat transfer mode changes from single-phase natural convection to nucleate pool boiling.[25, 58] We expand this notion to see whether the ONB boiling parameters can be used to shed light on mechanisms governing the HTC as well. Since the HTC is an ever-evolving parameter highly governed by bubble dynamics, we restrict the study scope to $\bar{h}_{low}$, where the bubble dynamics show main characteristics of the ONB. The side-view bubble evolution processes near the ONB of all samples are shown in **Figure 6**a-e, where tracked bubbles are overlayed with pixel-wise masks (See Movie S2 for real-time videos). Extracted bubble statistics shows that $f_b$ scales with $\sim D_b^{-1.5}$ (Figure 6f), which is a byproduct of the surface compensating for smaller bubbles by ejecting them at faster rates to balance heat removal. The error bars represent the 16$^{th}$ and 84$^{th}$ percentile of the data, meaning that 68% of the datapoints reside within the error bar range.

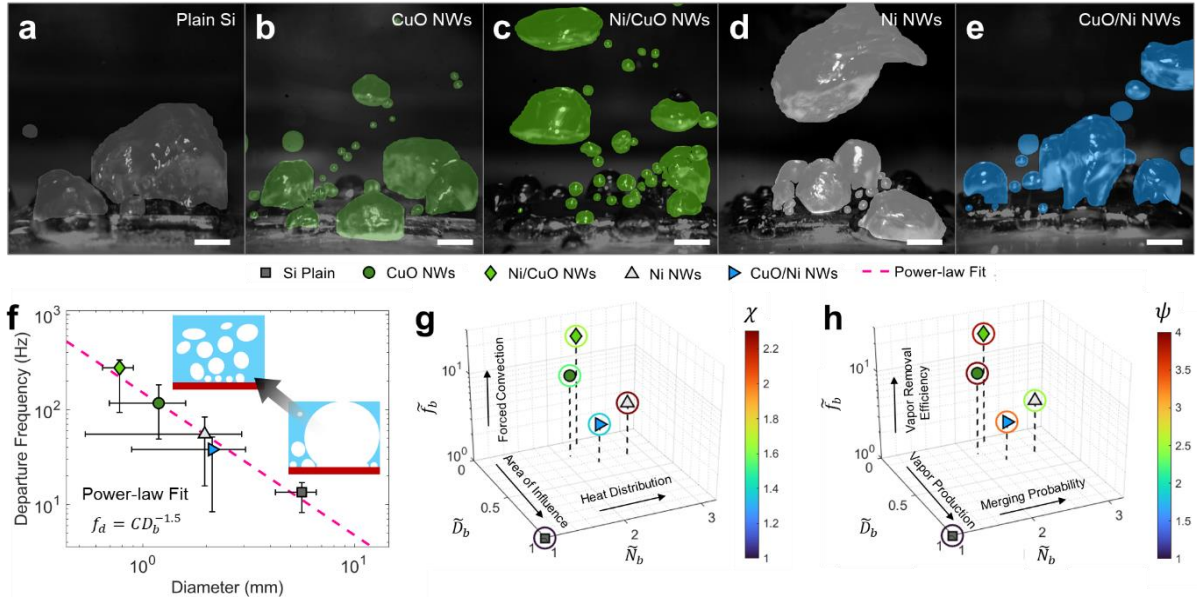

**Figure 6.** Bubble nucleation and departure behavior on unmodified and modified surfaces near the ONB. Representative side-view high-speed optical images of tracked bubbles on (a) plain Si (which is the unmodified surface), (b) CuO NW, (c) Ni/CuO, (d) Ni, and (e) CuO/Ni surfaces. The masks are color-coded with respect to the surface type. The scale bar is 2 mm. (f) Bubble departure frequency ($f_d$) as a function of departure bubble diameter ($D_b$) shows that that $f_d$ scales with $\sim D_b^{-1.5}$. The error bar represents the 16$^{th}$ and 84$^{th}$ percentile point, meaning that 68% of the datapoints lie within the error bar range. The inset illustrates how bubble characteristics change as a function of $f_d$ and $D_b$. Four-dimensional plot of the (g) HTC enhancement ratio ($\chi = \bar{h}_{low,modified}/\bar{h}_{low,unmodified}$) at low (<20 $Wcm^{-2}$) heat fluxes and (h) CHF enhancement ratio ($\psi = CHF_{modified}/CHF_{unmodified}$) as a function of bubble departure frequency ratio ($\tilde{f}_b = f_{b,modified}/f_{b,unmodified}$), departure diameter ratio ($\tilde{D}_b = D_{b,modified}/D_{b,unmodified}$), and effective nucleation site density ratio ($\tilde{N}_b = N_{b,modified}/N_{b,unmodified}$). The enhancements are represented with colored circles around the marker.

On the one hand, surfaces with wettable bases (i.e., Ni/CuO and CuO NW surfaces with CuO at the bottom) display orders of magnitude higher $f_b$ than the unmodified Si surface, demonstrating superior vapor removal capacities. On the other hand, surfaces with non-wettable bases (i.e., Ni and CuO/Ni NW with Ni at the bottom) display a wide $D_b$ distribution, suggesting a systemic failure of the surface to rid itself of small bubbles. This reiterates the importance of engineering wettabilities for optimal bubble activities. Despite having surface properties favorable for bubble nucleation, the retarded wetting of the hydrophobic Ni NWs hinders bubble departure when positioned at the sample base. Based on the acquired bubble statistics, we thus infer that $f_b$ is governed by the NW's base wettability, where bottom pressure fields generated by the out-of-plane wicking is speculated to help detach bubbles forming on the top.

The bare Si surface show the overall poorest bubble evolution and departure behaviors, with low $f_b$ and large, narrowly distributed $D_b$.

VISIONiT allows us to further quantify the multi-dimensional relationship between boiling parameters, heat transfer performances, and structural design, by defining non-dimensional HTC and CHF enhancement ratios compared to the unmodified Si, $\chi = \bar{h}_{low,modified}/\bar{h}_{low,unmodified}$ and $\psi = \text{CHF}_{modified}/\text{CHF}_{unmodified}$, respectively. Figure 6g and h shows four-dimensional (4D) maps of $\chi$ and $\psi$, with respect to departure diameter ratio ($\widetilde{D}_b = D_{b,modified}/D_{b,unmodified}$), departure frequency ratio ($\tilde{f}_b = f_{b,modified}/f_{b.unmodified}$), and nucleation site density ratio ($\widetilde{N}_b = N_{b,modified}/N_{b,unmodified}$), where $\chi$ and $\psi$ are represented as the fourth dimension with a color-mapped circle. The subscripts *modified* and *unmodified* indicate the nanostructures (nanowires in this study) and the bare silicon substrate. A summary of the boiling performance enhancements is provided in Table 2. From the boiling efficiency perspective (Figure 6g), $\chi$ (i.e., HTC ratios at low heat flux) mechanistically depends on liquid perturbation-induced forced convection at the wake of bubble departure (i.e., $\tilde{f}_b$), the area of influence (i.e., $\widetilde{D}_b$), and the heat transfer distribution along the surface (i.e., $\widetilde{N}_b$). The influence of $\tilde{f}_b$ is conspicuous in Figure 6g, where the $\chi$ of surfaces with similar $\widetilde{N}_b$ or $\widetilde{D}_b$ are ranked following the order of $\tilde{f}_b$. However, it becomes clear that $\widetilde{N}_b$ dominates HTC at low heat fluxes by observing that the Ni NW surface has the highest $\chi$ of ~2.3 despite having 80% lower $f_b$ than the Ni/CuO NW surface. The findings elucidate that at early boiling stages ($q'' < 20\ Wcm^{-2}$), it is necessary to design the boiling surface to dissipate heat evenly, rather than aggressively to maximize heat transfer.

From the boiling crisis standpoint (Figure 6h), $\psi$ relies on the quantity of vapor generation (i.e., $\widetilde{D}_b$), vapor removal efficiencies (i.e., $\tilde{f}_b$), and merging probabilities (i.e., $\widetilde{N}_b$). While the unmodified surface is primarily limited by vapor removal deficiencies and excessive vapor production, the NWs are restricted by a combination of $\widetilde{D}_b$, $\tilde{f}_b$ and $\widetilde{N}_b$. For example, the bubbles on the surfaces with non-wettable bases (i.e., Ni NW and CuO/Ni NW surfaces) tend to (i.e., non-wettable base) form with high nucleation densities, but depart with low frequencies. The tendency for a dense batch of bubbles with poor detachment capabilities to form naturally leads to increased bubble-bubble interactions, ultimately resulting in relatively large bubbles, which can potentially trigger the CHF. In contrast, the surfaces with wettable bases (i.e., CuO and Ni/CuO NW surfaces) proficiently remove vapor thermal barriers from the surface at reasonable nucleation densities and is hence assumed to be capable of sustaining higher heat loads.

## DISCUSSION

Achieving high HTC and CHF have conflicting requirements on the surface morphological and chemical properties. The CuO NW surface is an excellent example of how hierarchical structures with ultra-rough, superhydrophilic surface properties have been able to achieve high boiling performances in both aspects. Our work further pushes the limit of traditional hierarchical surface designs by combining two distinct surface properties into one nanoarchitecture, thereby tailoring the design for efficient vapor and liquid pathways during pool boiling. The liquid rate-of-rise tests confirm that the hierarchical NWs retain exceptional capillary performance (i.e., *K/R$_{eff}$*), even when combined with non-wicking Ni NW arrays. As a result, our hybrid Ni/CuO NW surface not only achieves 212% and 293% increases in HTC and CHF, respectively, when compared to an unmodified plain Si surface, but also improves the HTC at low-heat flux ranges by 9% and advances the ONB by 20% compared to the cutting-edge hierarchical CuO NW surface, while maintaining same levels of CHF performance.

Furthermore, the developed deep learning-based computer vision framework enables greater mechanistic understanding of the boiling process by connecting heat transfer and the large spatiotemporal bandwidth of bubble statistics. Here, the autonomous curation of massive *in-situ* bubble datasets enabled by our framework stands to compliment recent advances in imaging techniques, that exclusively study top- and bottom-view bubble statistics to identify the fundamental physics governing pool boiling.[53, 54, 59, 60] The datasets acquired at the ONB confirmed the correlations between bubble statistics and heat transfer performances. Due to the multidimensionality of the problem and innate structural heterogeneity of the hybrid NWs, still it is challenging to precisely predict overall HTC and CHF. Future research will include a systematically designed parametric study of bubble dynamics throughout the entire boiling curve.

In conclusion, our results confirm our hypothesis that boiling surfaces can be optimized for both high-and-low heat flux applications through a synergistic cooperation of hierarchical architecture and heterogeneous wetting properties. The combined efforts of surface chemistry, materials science, thermal engineering, and deep learning techniques shed light on the mechanistic relationship between vapor/liquid pathways, bubble statistics, and overall heat and mass transfer performance. Our proposed surface and characterization methods will help pave new design guidelines for other energy applications including boiling, condensation, thin-film evaporation, and microscopic liquid transport devices.

## EXPERIMENTAL SECTION

**Fabrication and Surface Modification**

Fabrication of the segmented nanowires is composed of three steps: 1) substrate preparation, 2) templated electrodeposition of metals, and 3) surface modifications. First, a Ti/Au ($5/50\ nm$) layer is deposited via evaporation on a $12\ mm \times 12\ mm$ Si subtrate. The Cu layer is additionally deposited onto the backside of the substrate using an E-beam evaporator. The nanowires are fabricated via electrodeposition using a sacrificial track-etched polycarbonate membrane. The membrane consists of pore diameters of approximately $1,000\ nm$. Metals (Cu and Ni) are then electrodeposited in an electrochemical cell containing electrolyte that is confined by an equally sized aperture. For deposition, a constant voltage of $V = -320$ mV (Ag/AgCl) and $V = -1100$ mV (Ag/AgCl) for Cu and Ni is applied, respectively. The electrolyte is composed of 0.6M $CuSO_4$ + 30mM $H_2SO_4$ for Cu deposition and 1M $NiSO_4$ + 0.2M $NiCl_2$ + 0.6M $H_3BO_3$ for Ni Deposition. The as-fabricated nanowires are further treated by selective surface oxidation after resolving the sacrificial template in Dichloromethane (Sigma Aldrich, >99V) for 1.5 hrs at 40ºC on the hot plate. The surface oxidation is treated by immersing nanowires in a preheated alkaline solution that is composed of a 1.7M $NaClO_2$ (Sigma Aldrich, > 99%) and a 0.25M NaOH (Sigma Aldrich, > 98% pellet) for 2 min. Finally, the nanowires are sequentially cleaned using deionized water, isopropyl alcohol, and acetone.

*Surface Characterization.* Morphological details of the nanowires are investigated using FEI Magellan 400 SEM at a 15 kV of accelerating voltage for both measurement from top view and cross-sectional view. To capture the cross-sectional images, the nanowires are cleaved and mounted on a vertical sample mount. Surface wetting properties are investigated via contact angle measurements using a sessile droplet method with a goniometer (MCA-3, Kyowa, Interface Sciences). The goniometer dispenses a 15 nL deionized water droplet from a 30 μm diameter capillary tip, assisted by constant air pressure (12 kPa). The droplet behavior during surface impact is captured by a side-view high-speed microscopic camera to measure contact angle (5,000 fps), and by a top-view optical microscope to record droplet infiltration and spreading dynamics (30 fps). The contact angle is evaluated by averaging results from three individual measurements in different surface locations.

**Liquid Rate-of-Rise Test**

The NW wicking performance is evaluated by the liquid rate-of-rise test (Figure S5), which calculates capillary performance parameters using the Lucas-Washburn equation. The liquid rate-of-rise test is performed in a saturated chamber with a water reservoir to prevent undesired evaporation during the test. The water reservoir is sealed with a parafilm for at least one week ahead of the test. The liquid rise is captured at 60 fps using a high-speed optical camera with a LED lighting source.

**Pool Boiling Test**

Pool boiling heat transfer of the nanowires is experimentally investigated using a custom-built pool boiling experimental setup (see Figure S8 and Supporting Information S3 for details). The pool boiling setup is composed of chamber, data acquisition system, copper block and heaters. The boiling chamber is composed of lid and transparent polycarbonate wall to capture the bubble dynamics using the high-speed camera (FASTCAM Mini AX50) at 2000 fps with LED lighting source. Heat flux and surface temperature in boiling test are calculated from the temperature results obtained from data acquisition system (LabJack U6) connected to four K-type thermocouples to read temperatures of the copper block. Thermal energy is generated from the four cylindrical cartridge heaters (Omega, CIR-20191) that are inserted at the bottom of the copper block. The heat flux is controlled by slowing increasing power using voltage transformer (Variac AC variable voltage converter). Before the boiling test, sample is placed and attached on the 1 cm × 1 cm copper block through soldering. Once the sample is firmly fixed on the copper block with proper PDMS sealing, the chamber is filled with working fluid (DI water). To maintain the water temperature close to saturation temperature, a guard heater that is connected to proportional integral derivative controller is used and immersed in the water.

*Pool Boiling Model Training and Evaluation.* We train a deep neural net object detector (i.e., Mask R-CNN) that generates pixel-wise masks for every object in an image. The object detection model trains on a custom-built image inventory of manually labeled bubble images gathered through years of experiments. To diversify the training data, we implement image augmentation techniques, which randomly transforms the original data into new, slightly modified versions. The total augmented training dataset (6944 images) are split into 80% (5555 images) train and 20% (1389 images) test sets. The model trains for a total of 100 epochs using stochastic gradient descent with a learning rate of 1e-3 and momentum of 0.9. The training results are presented in Figure S9b, showing that both train and test losses drop considerably low with a test loss of ~0.08, which indicates successful learning. Our model performs exceptionally well (> 90%) on traditional evaluation metrics such as recall, precision, accuracy, and F1-score. Furthermore, we report a pixel-wise mean average percentage error (MAPE) of 1.56% and an occlusion-induced error of approximately 4.6% (Figure S9c).

**ACKNOWLEDGEMENTS**

J.L. and Y.S. contributed equally to this work. This work was sponsored by the National Science Foundation (NSF) (Grant No. CBET-TTP 2045322, Thermal Transport Processes). J.L., and Y.S. are grateful for the financial support from the UC Irvine Mechanical and Aerospace Engineering Department Graduate Fellowship and support received through the collaborated work with Northrop Grumman Basic Research Center. Material characterizations were performed at the UC Irvine Materials Research Institute (IMRI).

Supporting Information

**Deep vision-inspired bubble dynamics on hybrid nanowires with dual wettability**

*Jonggyu Lee[1,§], Youngjoon Suh[1,§], Max Kuciej[2,3], Peter Simadiris[1], Michael T. Barako[3], Yoonjin Won[1,4,*]*

Co-First Authors[§], and Corresponding Author[*]

**This PDF file includes:**

    Section S1: Surface characterization
    Section S2: Pool boiling experimental setup
    Section S3: Bubble dynamics analysis using computer vision
    Figure S1. Cross-sectional SEM images of the nanowires
    Figure S2. Surface characterization of flat surfaces with replicated nanotextures
    Figure S3. Contact angle measurements of NW samples before and after oxidation
    Figure S4. Droplet imbibition and spreading during contact angle measurement of nanowires after oxidation
    Figure S5. Schematic illustration of experimental setup for liquid rate-of-rise test
    Figure S6. Liquid rise on Ni NWs at 10 s
    Figure S7. Liquid pathways through segmented Ni/CuO nanowire arrays
    Figure S8. Schematic of pool boiling experimental setup
    Figure S9. Computer vision framework and model performance evaluation.

    Table S1. Summary of extracted bubble parameters and heat flux values near the ONB.

    Supporting information references (1-12)

**Other Supporting Information for this manuscript include the following:**

    Movie S1. Real-time visualization of two-step wicking on a Ni/CuO NW surface.
    Movie S2. Real-time tracked bubble evolution behavior.

# Section S1. Surface characterization

## S1.1. Cross-sectional view of fabricated nanowires

Figure S1 shows cross-sectional morphological details of the segmented nanowires (NW) taken from a scanning electron microscope (SEM). To create NWs with dual wettability, two metals (Ni and Cu) are selected based on thermal conductivity, intrinsic contact angle, availability for electrochemical deposition, and corrosion resistance. As Ni resists corrosion, the Cu portion is selectively oxidized and decorated by CuO nanofeatures that provide enormous surface roughness and hydrophilicity, while the Ni portion remains intactly hydrophobic. The ratio of the Cu and Ni portion lengths ($l_{Cu}$, $l_{Ni}$) and average diameter are investigated by measuring and averaging the nanowire height and diameter in Figure S1.

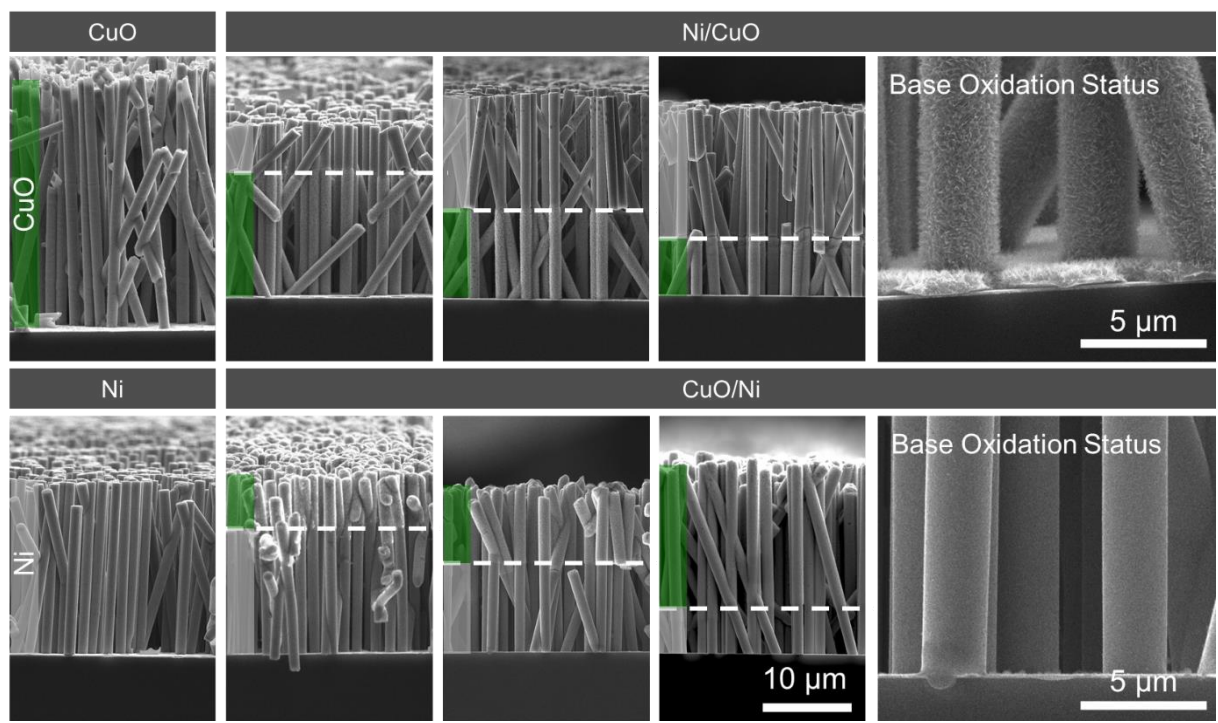

Figure S1. Cross-sectional SEM images of the nanowires.

**S1.2. Nanowire array porosity**

To quantify the porosity distribution of the fabricated samples, we calculate the effective porosity of the NW by accounting for the effects of the stepwise diameter increase in the Cu segment after oxidation. The effective porosity is defined as the void fraction of the system $\phi = 1 - \varepsilon$, where $\varepsilon$ is the NW solid fraction. The solid fraction, in turn, is calculated as $\varepsilon = \pi d^2 N/4$, where $N$ is the NW areal number density.[1] The number density of nanowires is $N$=0.23 μm$^{-2}$ obtained from the top-view SEM images for all the nanowires in this study. Top-view SEM image comparison of the CuO NWs (Figure 2i) and Ni NWs (Figure 2j) confirm that the chemical oxidation has no noticeable effects on the NW's natural tortuosity. Therefore, the overall porosity of segmented nanowires is impacted by the ratio of the segmented Cu and Ni portions, owing to the difference in diameter between the two metal segments. Considering all these factors, the porosity of the segmented nanowire after oxidation can be calculated as follows:

$$\phi = 1 - \frac{\pi N}{4 l_{tot}} (d_{Ni}^2 l_{Ni} + d_{CuO}^2 l_{CuO}) \tag{S1}$$

The calculation results are listed in the Table 1. We report an insignificant (<10%) porosity discrepancy across all NW samples used for this study. Note that the areal loss from the entanglement of the nanowires is evaluated as 0.2% from our previous study and is factored into our porosity evaluation in this study.[1]

## S1.3. Surface chemistry and wetting behavior

The NW's surface chemistry changes due to the selective oxidation of the Cu material. To directly compare the coupled effects of NW structures and oxidation, we first prepare plain Ni and Cu surfaces using identical substrates as shown in Figure S2a. After oxidation, the contact angle remains relatively stable ($\Delta\theta < 8\%$) for the Ni surface while noticeably dropping by $\Delta\theta = 37\%$ for the Cu surface (Figure S2b). The oxidation-induced contact angle drop becomes more prominent when coupled with NW geometry. By comparing Figure S2 and Figure S3, it becomes evident that the contact angles of the NW surfaces before oxidation are generally higher than the plain element surfaces owing to the additional roughness introduced by the NW arrays.

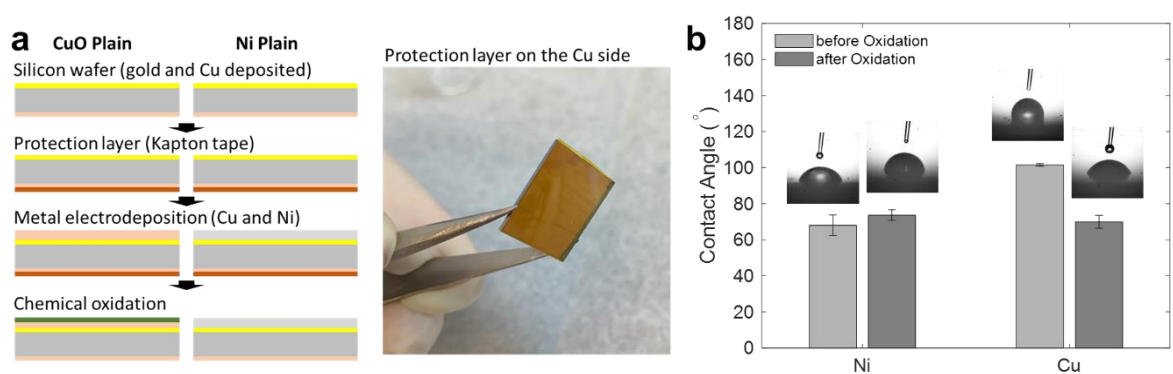

Figure S2. Surface characterization of flat surfaces with replicated nanotextures. (a) Sample preparation of metal surfaces coated on both bottom and top of the silicon wafer. The plain and structured metal surfaces are fabricated on an Au layer using an electrodeposition method. A Cu layer is coated via E-beam evaporation on the backside of the sample to provide a metallic surface required for the soldering process. (b) Contact angle results of the plain metal surface before and after oxidation.

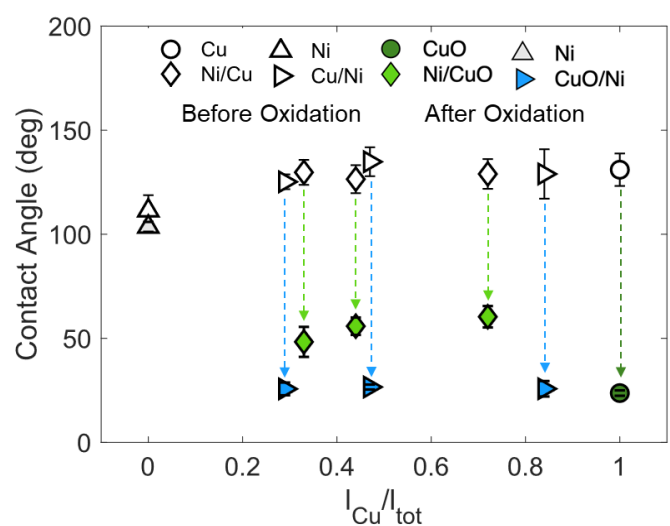

Figure S3. Contact angle measurements of NW samples before and after oxidation.

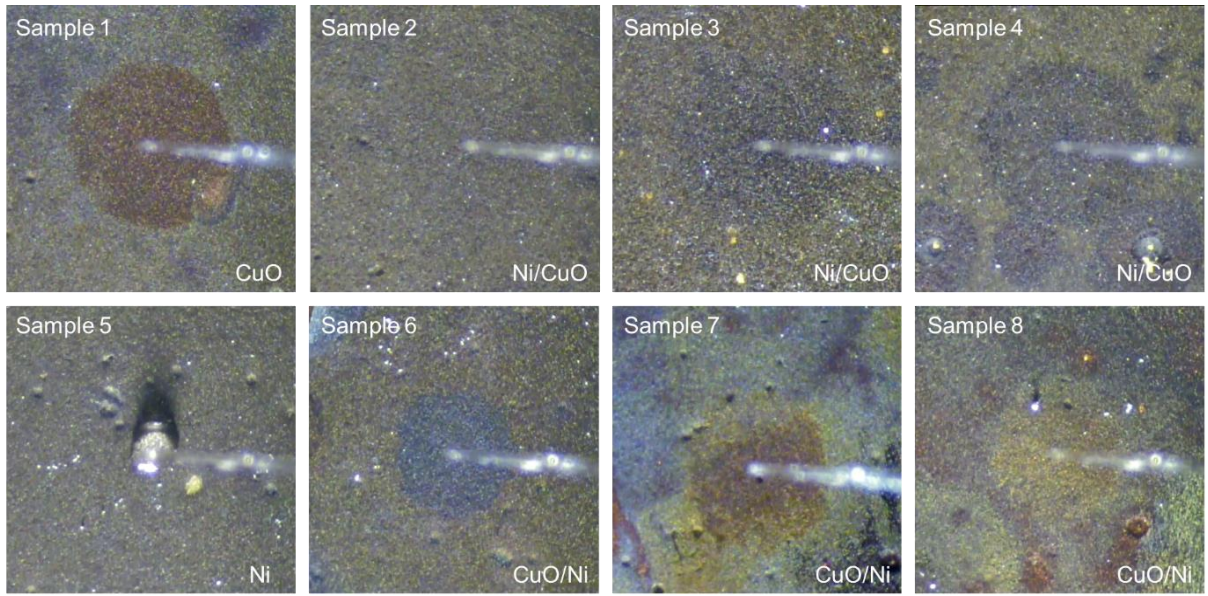

Figure S4. Droplet imbibition and spreading during contact angle measurement of nanowires after oxidation.

**S1.4. Capillary rise theory**

The capillary pressure through the porous media is dictated by the surface tension of the liquid and interfacial interaction between liquid-vapor. The relation of the capillary pressure $P_{cap}$ can be expressed with the Laplace-Young equation:

$$\Delta P_{cap} = \frac{2\sigma \cos \theta_s}{R_p} \quad (S2)$$

where $\sigma$ is the surface tension of the liquid, and $R_p$ is the static pore radius. During the wicking process, the capillary pressure should be equal to the summation of the viscous friction, gravity, and evaporation for momentum balance as given by:

$$\frac{2\sigma \cos \theta_s}{R_p} = \frac{\phi}{K}\mu h v + \frac{\dot{m}_{evp}\mu}{2d_{film}\rho K}h^2 + \rho g h \quad (S3)$$

where $K$ is permeability, $\mu$ is the liquid viscosity, $d_{film}$ is film thickness, and $\rho$ is the liquid density. The mass evaporation rate $\dot{m}_{evp}$ is assumed to be zero in the vapor saturation condition in the chamber. The gravitational term including the gravitational acceleration $g$ is negligible in the condition of such a low wicking height (< 10 mm) at the permeability over $1\times10^{-20}$ m². Therefore, the Equation S2 can be reduced to the Lucas-Washburn equation[3,4]:

$$h^2 = \frac{4\sigma}{\phi\mu}\frac{K}{R_{eff}}t \quad (S4)$$

where the effective radius $R_{eff} = R_p/\cos\theta_s$.[1]

**S1.5. Uncertainty analysis for liquid rate-of-rise test**

*Porosity.* The porosity is calculated by the number density *N* and diameter *d* of the NW arrays. The number density is calculated by counting the number of NWs within a designated window frame (15 µm × 15 µm). In order to minimize the uncertainty associated with the NWs, we count >200 instances for all sets. The uncertainty in the diameter is identified based on the diameter distribution of each sample group. Therefore, the uncertainty in the porosity is

$$\left(\frac{\Delta\phi}{\phi}\right)^2 = \left(2\frac{\Delta d}{d}\right)^2 + \left(\frac{\Delta N}{N}\right)^2 \tag{S5}$$

*Capillary performance parameter.* Based on the Equation S3, the capillary performance parameter is a function of porosity, fluid viscosity, surface tension, wicking height, and measurement time. The uncertainty in the capillary performance parameter is given by

$$\left(\frac{\Delta K/R_{eff}}{K/R_{eff}}\right)^2 = \left(\frac{\Delta\sigma}{\sigma}\right)^2 + \left(\frac{\Delta\phi}{\phi}\right)^2 + \left(\frac{\Delta\mu}{\mu}\right)^2 + \left(2\frac{\Delta h}{h}\right)^2 + \left(\frac{\Delta t}{t}\right)^2 \tag{S6}$$

The fluid properties vary with the lab temperature 298 ± 2 K; therefore, the uncertainty in the surface tension and viscosity is 0.6% and 4.4%, respectively, which can be negligible. Also, the framerate of the camera is 60 fps, leading to an uncertainty of approximately 8.4 $ms$, which is considered negligible (< 0.015%) over long measurement period. Therefore, the measurement error associated with wicking height $\Delta h$ is used to calculate the overall uncertainty.[1]

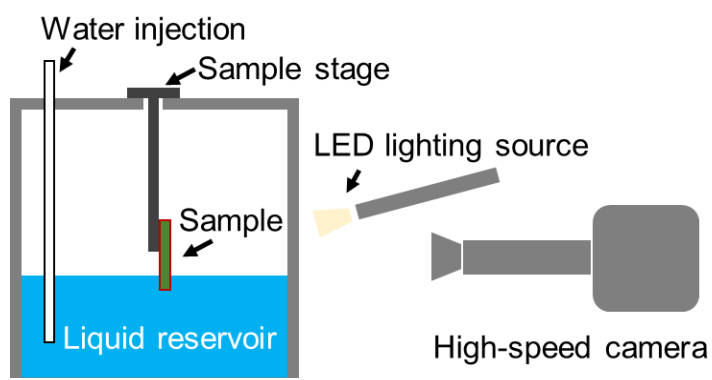

Figure S5. Schematic illustration of experimental setup for liquid rate-of-rise test. The setup is composed of a liquid chamber, water injection system, sample stage, and CCD camera. The water is injected gradually to control water height to initiate capillary rise. The test is performed in a saturation chamber to prevent evaporation during the liquid rise.

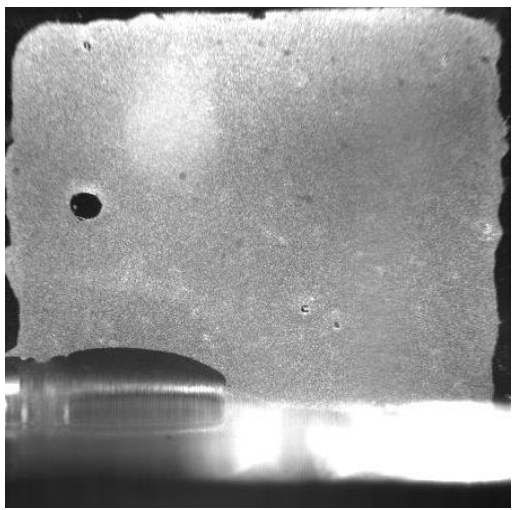

Figure S6. Liquid rise on Ni NWs at 10 s. Capillary wicking is not observed on the Ni NWs due to the hydrophobic property. The meniscus that is formed partially on the edge of the sample might be due to the exposure of the hydrophilic Au bottom layer.

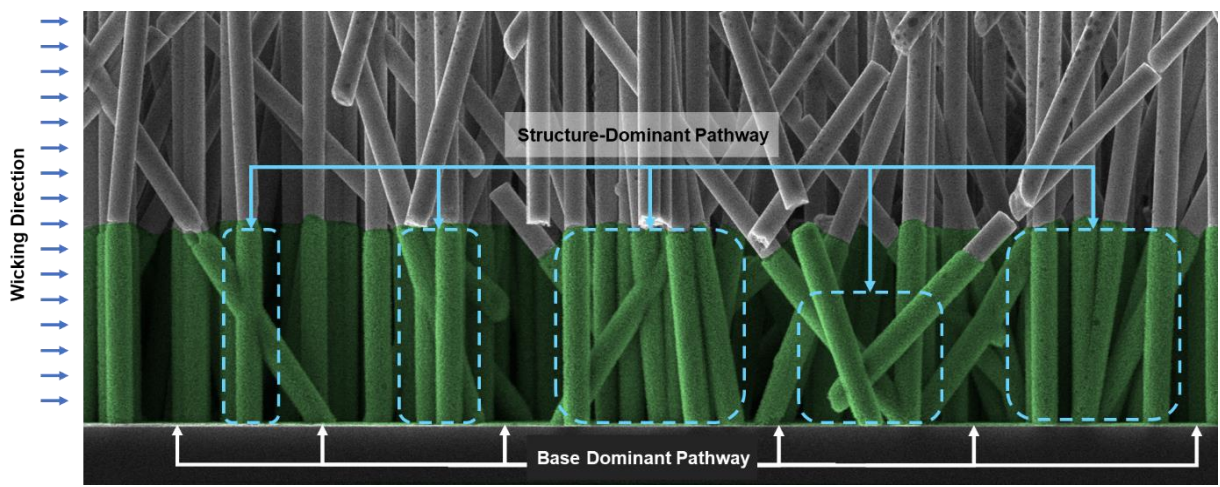

Figure S7. Liquid pathways through segmented Ni/CuO nanowire arrays. Liquid travels in the path of least hydraulic resistance. Structure-dominant liquid pathways form in regions where NW arrays are clustered due to the reduction of effective wicking distances. When NWs are spaced far apart, the liquid primarily wicks through the hydrophilic CuO base layer. Therefore, at low CuO content, the wicking is promoted through the efficient wicking pathways through the base layer. The capillary performance reduces with increasing CuO content as structure-dominant pathways start governing the overall liquid delivery.

## Section S2. Pool boiling experimental setup

### S2.1. Uncertainty analysis for pool boiling experiment

The uncertainties for pool boiling experiments are computed by using the law of propagation of uncertainty. The heat flux $q'' = k\Delta T/L$ is a function of temperature gradients, material properties, and thermocouple positions. $q''$ is calculated by averaging the measured $q''$ values obtained from thermocouples 1 – 4 as shown:

$$q'' = k\left[\frac{\left(\frac{T_1 - T_2}{L_1}\right) + \left(\frac{T_2 - T_3}{L_2}\right) + \left(\frac{T_3 - T_4}{L_3}\right)}{3}\right] \tag{S7}$$

where $T_{i=1,2,3,4}$ are the temperature readings from the four thermocouples used in the experiment, $k$ is the thermal conductivity, and $L_{i=1,2,3}$ are the distance between thermocouples.

By assuming that the thermal conductivity remains constant during experiments and that positional errors are minimized, the uncertainties become dictated by thermoucple readings ($U_T = \pm 1.1$°C). As a result, the uncertainty of the heat flux becomes:

$$U_{q''} = \sqrt{\left(\frac{\partial q''}{\partial T_1}U_T\right)^2 + \left(\frac{\partial q''}{\partial T_2}U_T\right)^2 + \left(\frac{\partial q''}{\partial T_3}U_T\right)^2 + \left(\frac{\partial q''}{\partial T_4}U_T\right)^2} \tag{S8}$$

By solving for Equation (S8), an uncertainty of approximately 2% is calculated for the maximum heat flux.

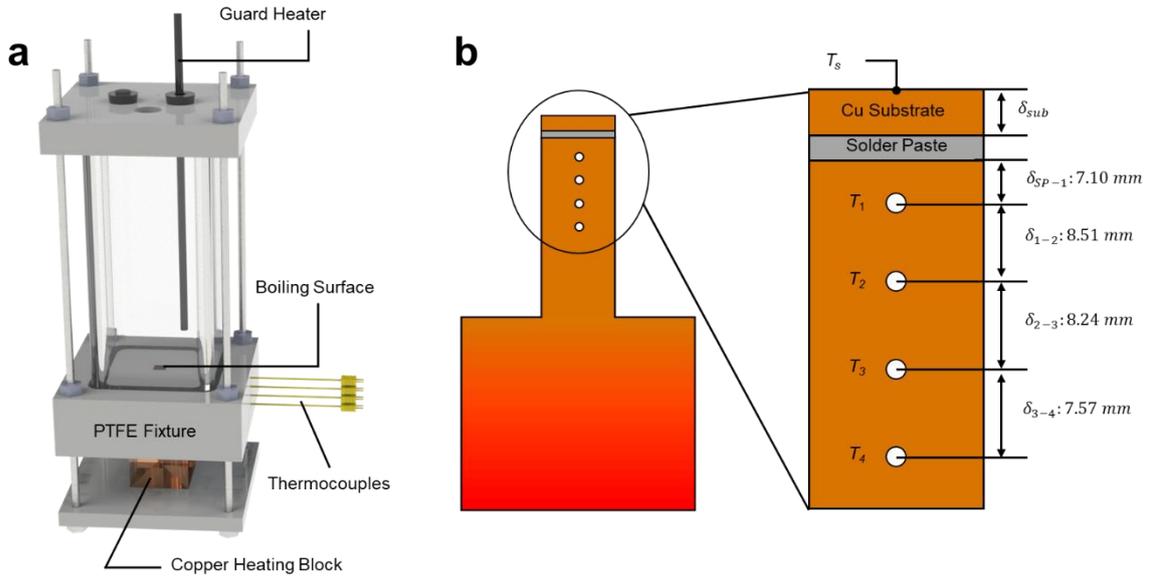

Figure S8. Schematic of pool boiling experimental setup. (a) The pool boiling setup consists of a clear boiling chamber, PTFE fixture, a guard heater, and thermocouple-embedded copper heating block. (b) The samples are soldered onto the boiling surface as shown. The heat flux is estimated by averaging the measurements obtained from four embedded thermocouples with illustrated spacings.

**Section S3. Bubble dynamics analysis using deep learning**

**S3.1. Intelligent vision-based framework**

Our developed framework consists of an object detection, object tracking, and data processing module to extract physical descriptors from experimentally acquired pool boiling image datsets. 1) Object detection module: As illustrated in Fig. S9a, high-resolution droplet images first pass through a custom-trained object detection module (Mask R-CNN) where droplet masks assigned with unique IDs are generated. At this stage, the model records primitive spatial features (e.g., equivalent diameter, pixel-wise area, eccentricity, orientation, solidity, and location). 2) Object tracking module: The detected masks then pass through a tracking module (TrackPy) where the IDed spatial features are used as parameters for tracking via the k-dimensional (k-d) tree algorithm. During the bubble tracking process, potential errors are manually identified and corrected using a documented graphical user interface (GUI). The model accuracy from the object detection/tracking is validated by testing evaluation metrics such as recall, precision, accuracy, F1-score, mean average pixel error (MAPE), and occlusion-induced errors, see Section S3.2 for details. 3) Data processing module: The datasets are finally post-processed to extract higher-level features (e.g., departure bubble diameter $D_b$, departure frequency $f_b$, and effective nucleation site density $N_b$) and visualized via Matlab.

**S3.2. Model evaluation**

To validate our model prediction performance, we develop a MATLAB script that checks if the predicted dataset pixels correspond with the labeled data (i.e., ground truth (GT)). A positive condition is when a model detects an instance that matches the GT and is counted as a true positive (TP). By contrast, a false positive (FP) instance is counted when the model predicts an nonexisting object. Simlarily, true negative (TN) situates when the model correctly predicts no existing instances, and false negative (FN) counts instances when an existing object is undetected. These conditions are summed across the dataset and used to determine the standard performance metrics of an object detection model:[2, 3]

$$Accuracy = \frac{TP + TN}{TP + TN + FN + FP} \quad (S9)$$

$$Recall = \frac{TP}{TP + FN} \quad (S10)$$

$$Precision = \frac{TP}{TP + FP} \quad (S11)$$

$$F1\ Score = 2 * \frac{(Precision * Recall)}{(Precision + Recall)} \quad (S12)$$

Furthermore, we define a new performance metrics called the mean average percentage error (MAPE) for more detailed evaluations of the model prediction accuracies at the pixel-level. For this, a pixel-wise error (PE) is calculated by subtracting the ground truth binary mask from the predicted binary mask (PBM), then dividing by the ground truth. This results in the true negatives being removed from the binary matrix, leaving only true positive, false positive, and false negative pixels. MAPE is then calculated as:

$$MAPE = \frac{1}{n}\sum_{i=1}^{n}|PE| \times 100 = \frac{1}{n}\sum_{i=1}^{n}\left|\frac{(GT - PBM)}{GT}\right| \times 100 \qquad (S13)$$

Our framework displays striking performance (>90%) on all metrics (Figure S9b). To characterize occlusion-induced errors, we manually compare a dataset consisting of >200 random labelled images with model predictions and solely estimate occlusion-induced errors of surface bubbles by assuming spherical morphologies. We report a maximum ocllusion-induced error of 4.6%.

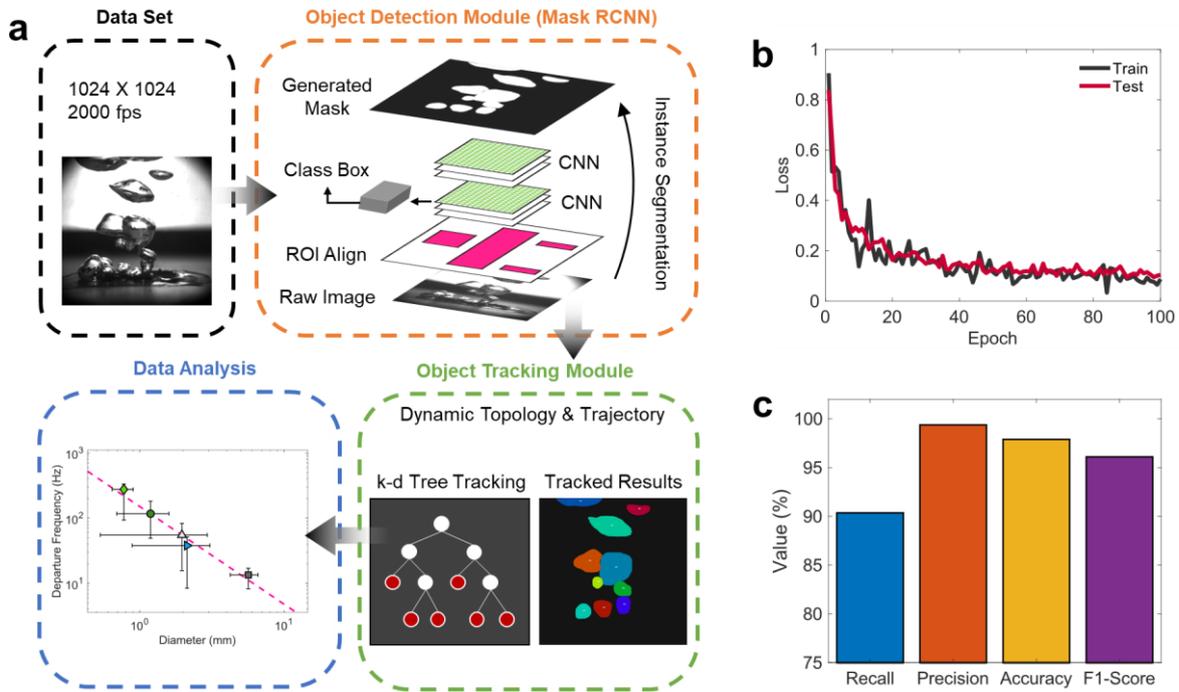

Figure S9. VISIONiT framework and model performance evaluation. (a) The object detection module uses high-speed, high-resolution images as an input dataset. Next, the images are passed through an object detection module (Mask R-CNN) where bubbles are automatically detected and labelled with pixel-wise masks. The masks are then linked together with respect to time through k-dimensional tree algorithms in the tracking module. Finally, the tracked results are post-processed in the data analysis module for visualization. (b) Mask R-CNN model learning curves show an exponential decaying trend. (c) The model performs exceptionally well on all traditional object detection evaluation metrics.

### S3.2. Mechanistic pool boiling models and detection algorithms

The boiling heat flux released from the surface $q_{tot}^"$ is partitioned by the three principle components, namely, natural convection $q_{nc}^"$ outside of the bubble-influenced domain (i.e., area of influence $A_{inf}$), transient conduction $q_{fc}^"$ over the area of influence, and evaporative heat transfer $q_{ev}^"$ during phase change as follows:[4-12]

$$q_{tot}^" = q_{nc}^" + q_{fc}^" + q_{ev}^" \tag{S14}$$

$$q_{nc}^" = \left(1 - N_b \frac{\pi D_b^2}{4}\right) h_c \Delta T \tag{S15}$$

$$q_{fc}^" = \frac{1}{2}\left[D_b^2 N_b \left(\sqrt{\pi k \rho c f_b}\right)\right] \Delta T \tag{S16}$$

$$q_{ev}^" = N_b f_b \left(\frac{\pi}{6} D_b^3\right) \rho_v h_{fg} \tag{S17}$$

where $N_b$ is the effective nucleation site density, $D_b$ is the departure diameter, $h_c$ is the average convective heat transfer coefficient (HTC) outside of the $A_{inf}$, $\Delta T$ is the superheat, and $\rho$ is the density of the liquid, $c$ is the heat capacity, $f_b$ is the departure frequency and $h_{fg}$ is the latent heat of evaporation. We note that $h_c$ remains experimentally undefined because the convective heat transfer varies depending on the surrounding bubble nucleation and departure behaviors. Therefore, we suspect that $h_c$ will be a function of the experimentally measured HTC and assume $h_c \approx 0.5 h_{exp}$, where $h_{exp}$ is the experimentally measured HTC.

In order to characterize the key boiling parameters, we develop custom algorithms to process spatiotemporal features with respect to the boiling surface. Bubble departure (BD) events are detected by comparing the bottom bounding box's relative position to the surface: $BD = IF(bbox_{bot,1} < y) \text{ AND } IF(bbox_{bot,2} > y)$, where $bbox_{bot,t}$ is the bottom bounding box coordinate at time $t$ and $y$ is the pixel value corresponding to the boiling surface. The departure frequency $f_b = 1/t_g$ is measured as a function of the bubble growth time $t_g$, which characterizes the time between bubble nucleation and departure. While classical theories include an additional time period, namely the waiting time $t_w$, which characterizes the time between bubble nucleation and departure, optical measurements confirm that $t_w \approx 0$ even at ONB. Please see Movie S2 for semi real-time mask predictions for all surfaces, where bubbles form immediately after departure. Lastly, the effective nucleation site density is estimated as $N_b = \left(\frac{1}{Z}\sum_{j=1}^{Z} x_j\right)/A_s$, where $x$ is the number of bubbles on the surface at time $j$, $Z$ is the total number of timesteps, and $A_s$ is the projected boiling surface area. To test our model's prediction validity, we compare the total theoretical heat flux $q_{th}^"$ calculated from heat partitioning analysis Equation S14-17 with experimental heat flux $q_{exp}^"$ values (Table S1), showing reasonable mean error of ~13%.

Table S1. Summary of extracted bubble parameters and heat flux values near the ONB.

| Surface Type | $D_b$ (mm) | $N_b$ (cm$^{-2}$) | $f_b$ (Hz) | $q''_{exp}$ (Wcm$^{-2}$) | $q''_{th}$ (Wcm$^{-2}$) |
|---|---|---|---|---|---|
| Si Plain | 5.64 | 2.1 | 13.45 | 10.32 | 11.23 |
| CuO NWs | 1.18 | 4.48 | 116.79 | 5.9 | 4.32 |
| Ni/CuO NWs | 0.77 | 4.79 | 275.45 | 3.75 | 3.74 |
| Ni NWs | 1.95 | 5.84 | 55 | 12.11 | 8.41 |
| CuO/Ni NWs | 2.12 | 5 | 37.95 | 4.87 | 4.93 |